# Silicon-Vacancy Color Centers in Nanodiamonds: Cathodoluminescence Imaging Marker in the Near Infrared

H. Zhang[1], I Aharonovich[2,6], D.R. Glenn[3], R. Schalek[4,5], A. P. Magyar[2],

J.W. Lichtman[4,5], E. L. Hu[2], R. L. Walsworth[1,3,5*]

[1]Department of Physics, Harvard University, Cambridge, MA, 02138

[2]School of Engineering and Applied Sciences, Harvard University, Cambridge, MA, 02138

[3]Harvard-Smithsonian Center for Astrophysics, Cambridge, MA, 02138

[4]Department of Molecular and Cellular Biology, Harvard University, Cambridge, MA, 02138

[5]Center for Brain Science, Harvard University, Cambridge, MA, 02138

[6]School of Physics and Advanced Materials, University of Technology Sydney,

Broadway, NSW, 2007, Australia

*rwalsworth@cfa.harvard.edu

## Abstract

*We demonstrate that nanodiamonds fabricated to incorporate silicon-vacancy (Si-V) color centers provide bright, spectrally narrow, and stable cathodoluminescence (CL) in the near-infrared. Si-V color centers containing nanodiamonds are promising as non-bleaching optical markers for correlated CL and secondary electron microscopy, including applications to nanoscale bioimaging.*

Imaging of biological structures and processes with nanoscale spatial resolution is an important goal at the interface of the physical and life sciences [1][2]. The recent development of novel super-resolution imaging techniques coincides with efforts to fabricate or synthesize optically bright and stable biomarkers [3]. In particular, nanodiamonds have emerged as promising candidates for nanoscale imaging and sensing. Nanodiamonds compare favorably with other synthetic nanoscale biomarkers due to: (i) their excellent biocompatibility; and (ii) their ability to host bright color centers and defects [4] that span most of the optical spectrum and can serve as precision sensors of magnetic fields [5][6][7], electric fields [8], temperature [9], and the chemical environment [10]. Optical fluorescence imaging and particle tracking [11], as well as nanoscale magnetic [12] and thermal [9] sensing have recently been demonstrated with nanodiamonds in living cells. Nanodiamonds containing nitrogen-vacancy (NV) centers have been localized at nanometer scales using stimulated emission depletion (STED) microscopy [13], as well as wide-field switchable emitter super-resolution imaging [14]

We recently demonstrated [15] multi-color cathodoluminescence (CL) of nanodiamonds as a powerful tool for nanoscale imaging of biological structures. CL is the emission of light by matter as the result of electron bombardment. CL imaging of bulk matter is typically carried out in an electron microscope outfitted with an optical detector, and is widely used in materials characterization [16]. However, application of CL to imaging biological structures has been hindered by low photon count rates and rapid signal degradation due to the destruction of biomolecules and organic fluorophores under electron beam irradiation [17]. These problems may be overcome with correlated CL and secondary electron (SE) imaging of samples tagged with surface-functionalizable nanoparticles containing defects that are robust under electron beam illumination and emit stable, spectrally distinct CL: e.g., A-band defects and NV centers in nanodiamonds, as well as Ce:LuAG nanophosphors [15]. In this approach, the CL-emitting particles function as color-distinguishable nanoscale markers of targeted epitopes, while the correlated SE image provides high-resolution information about the cellular structure. For applications with large intrinsic CL background, e.g., correlated CL and SE imaging of unfixed/living cells in an environmental chamber in a scanning electron microscope (SEM) [18], it is desirable to use spectrally narrow CL markers with emission peaks at wavelengths distinct from the CL background (such as from proteins, nucleic acid, and fluorophore-conjugated antibodies, which usually emit CL at short optical wavelengths) [19]. However, many of the nanoparticle species investigated to date have relatively broad (~100 nm) CL emission spectra at room temperature. Here, we show that silicon-vacancy (Si-V) color centers [20] in nanodiamonds provide a promising solution to this challenge.

Specifically, we demonstrate experimentally that nanodiamonds fabricated to incorporate Si-V color centers provide bright, spectrally narrow (~5 nm) CL emission in the near-infrared (~740 nm), which lies within the near-infrared transmission window of biological tissue and is thus well suited for bioimaging. Furthermore, since diamond is a wide bandgap material, with most optically active defects emitting in the ultraviolet and visible [21], an important advantage of Si-V CL is its spectral purity: i.e., the low intrinsic CL background from diamond in the near-infrared.

The Si-V color center consists of a single silicon atom adjacent to two vacancies in the diamond lattice [22]. Si-V centers can be engineered into nanodiamonds either by the addition of Si gas during diamond growth by chemical vapor deposition (CVD) [23][24], or by implantation of silicon ions into pristine diamond [20]. Si-V centers exist stably in either the negative charge state (Si-V)$^-$ with zero-phonon-line (ZPL) optical emission wavelength ≈738 nm, or the neutral charge state (Si-V)$^0$ with ZPL wavelength ≈947 nm. Current interest is primarily focused on the negative charge state (Si-V)$^-$ center because of the more efficient optics and detectors at its emission wavelength. Recently, the (Si-V)$^-$ color center has attracted attention from the nano-photonics and quantum information communities [25][26][27], as it is one of the brightest known solid-state single photon sources, emitting up to 4.8 x 10$^6$ photons/sec at room temperature [28]. The combination of high oscillator strength, narrow linewidth, and a relatively short excited electronic state lifetime (~1 ns) also make Si-V doped nanodiamonds promising candidates for use as CL biomarkers. Si-V CL was first observed in bulk polycrystalline diamond in 1980 [29], but there are no published studies of Si-V CL in nanodiamonds prior to the present work.

We investigated the material and CL properties of Si-V nanodiamonds synthesized by thin diamond film growth using plasma-enhanced chemical vapor deposition (PECVD), followed by sonic disintegration to generate isolated free particles. (See Experimental Section for details.) As a reference, we compared measurements of CL and photoluminescence (PL) from Si-V nanodiamonds, both for the negative and neutral charge states, to the well-understood, bright, and narrow-linewidth CL and PL emission at ~610 nm from rare-earth-doped Eu:$Y_2O_3$ nanophosphors, which are commonly used for field emission displays [30] and narrow-linewidth PL [31].

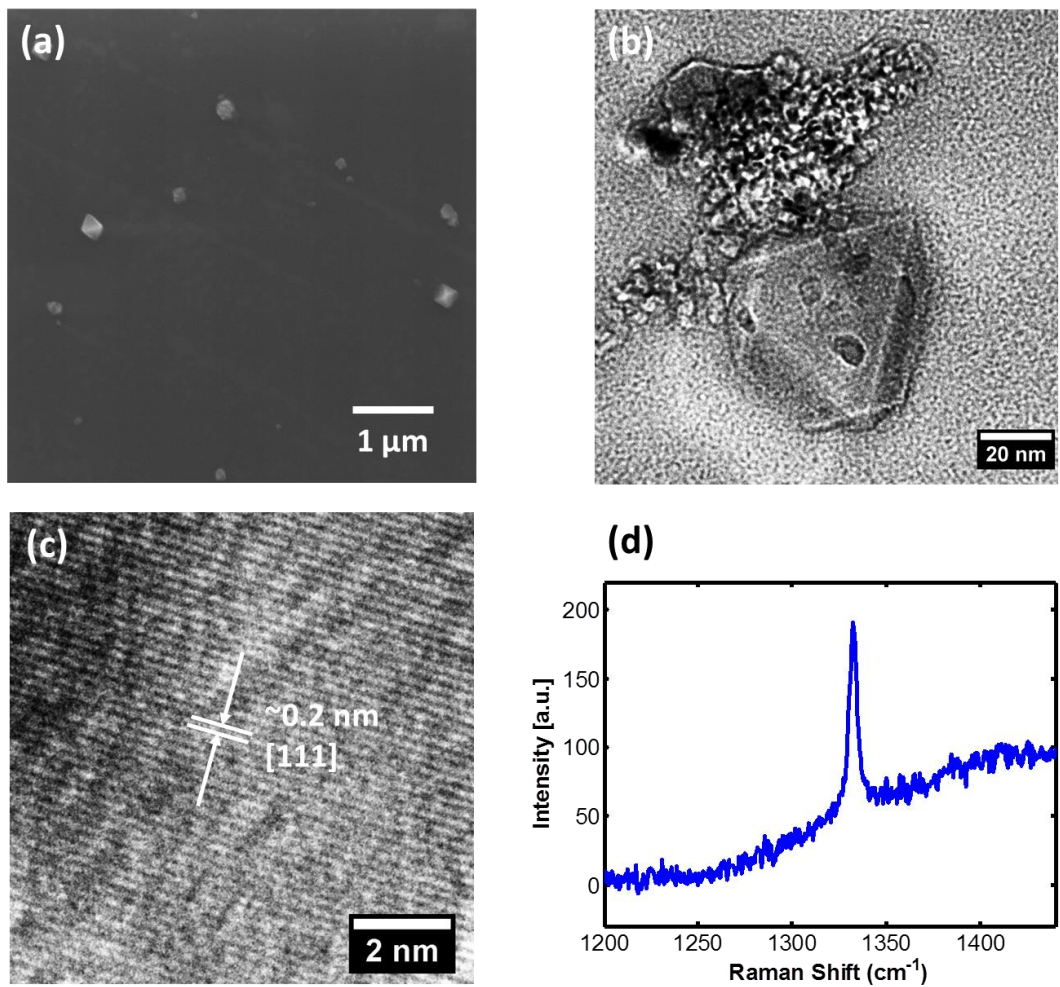

*Figure 1. Nanodiamonds containing Si-V defects were prepared by plasma-enhanced chemical vapor deposition (PECVD). (a) SEM image of nanodiamonds (~20-200 nm) grown on top of silicon substrate. (b) TEM image of ~ 40 nm nanodiamond shows diamond facet. Nanodiamonds were ultrasonically disintegrated from silicon substrate and dropcast onto TEM grid. (c) High resolution TEM image of same sample shows diamond crystal lattice. (d) Optical Raman spectroscopy of nanodiamonds exhibits diamond (i.e., $sp^3$-bonded carbon) peak.*

We characterized the size and material properties of these particles using electron microscopy (EM) and optical Raman spectroscopy, demonstrating the presence of a well-formed diamond lattice with minimal graphitization. **Figure 1a** shows a characteristic SEM image of Si-V nanodiamonds on a silicon substrate, with diamond facets clearly observable for most particles. Samples were recovered from the substrate by sonication, and dropcast onto a grid for transmission electron microscopy (TEM). A TEM image of a well-faceted PECVD-grown nanodiamond is shown in **Figure 1b**, sitting adjacent to much smaller nanodiamond seeds. High-resolution TEM indicates a lattice-spacing of 0.2 nm (see **Figure 1c**), which matches well with the expected value for {111} planes in diamond [32]. **Figure 1d** shows an optical Raman spectrum recorded from Si-V nanodiamonds at room temperature, with an excitation wavelength of 532 nm. A single sharp peak at 1332 cm$^{-1}$ corresponds to the first-order Raman-active phonon mode in diamond. The observed Raman FWHM linewidth of 5 cm$^{-1}$ is only slightly broader than the ~3 cm$^{-1}$ linewidth typically seen in bulk diamond in room temperature, indicating a high degree of crystallinity in the Si-V nanodiamonds [24]. Additional broad Raman resonances observed at higher wavenumbers suggest the presence of phases containing sp$^2$-bonded carbon atoms, which are frequently seen at bulk CVD diamond surfaces [33]. Surface oxidation [34] and acid cleaning are established procedures for removing these residual graphitic structures.

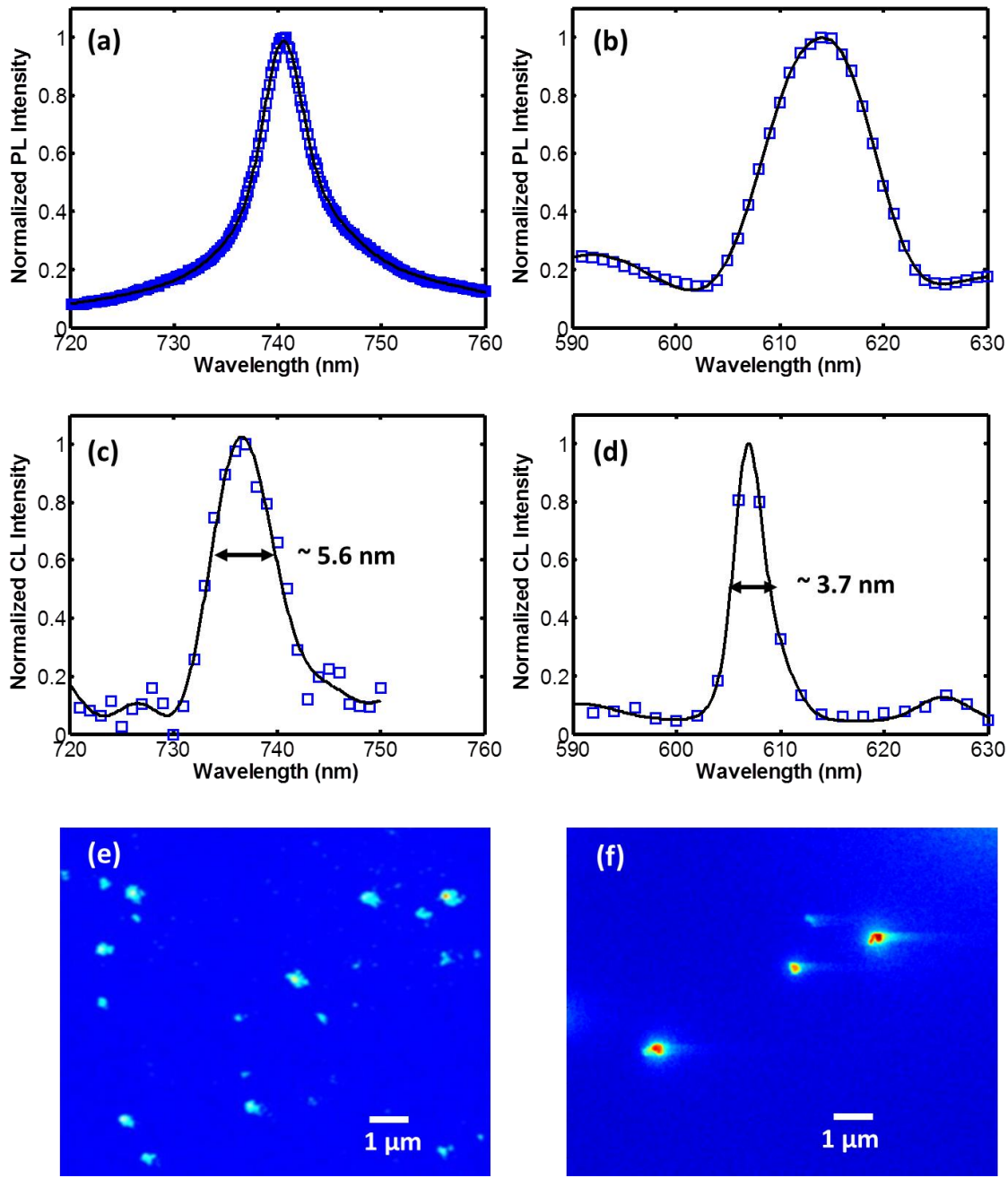

*Figure 2. Optical property comparison of nanodiamonds containing Si-V color centers and Eu:Y$_2$O$_3$ nanophosphors. (a) and (b) are PL spectra for Si-V nanodiamonds and Eu:Y$_2$O$_3$ nanophosphors, respectively. Si-V nanodiamond sample was excited by 532 nm laser, whereas Eu:Y$_2$O$_3$ nanophosphor, sample was excited at ≈230 nm by a spectrally filtered lamp inside a plate reader. (c) and (d) are corresponding CL spectra. Open squares represent experimental data and solid lines are fits to sums of Gaussians. (e) and (f) are CL images of Si-V nanodiamonds and Eu:Y$_2$O$_3$ nanophosphors, respectively. Since Si-V color centers have a much shorter lifetime than Eu:Y$_2$O$_3$, 2 ns [27] versus 2.8 ms [35], there is minimal image distortion in (e), but a long emission tail and hence significant image distortion in (f).*

We probed the optical properties of Si-V nanodiamonds using both PL and CL. A typical room-temperature PL emission spectrum under 532 nm excitation is displayed in **Figure 2a**, showing the characteristic narrow Si-V zero phonon line (ZPL) at 738 nm. A homebuilt, fiber-based light collection system and commercial grating spectrometer were used to obtain CL emission spectra from the same nanodiamond sample, as shown in **Figure 2c**. The main CL spectral feature is at approximately the same wavelength as the optical PL Si-V ZPL, and with comparable linewidth. For comparison, **Figures 2b** and **2d** present PL and CL spectra, respectively, from the transition at ~610 nm in Eu:Y$_2$O$_3$ nanophosphors, under 230 nm optical excitation (spectrally filtered lamp inside plate reader) for PL and electron beam excitation for CL. While the spectral width of the nanophosphor emission is comparable to that from Si-V nanodiamonds, the central wavelength of the Si-V nanodiamond emission coincides better with the near-infrared transmission window of biological tissue.

The short (~2 ns [27]) excitation lifetime of the Si-V color center, compared to nanophosphors, is also advantageous for CL-based imaging in an SEM, as illustrated in **Figures 2e** and **2f**. For typical electron beam parameters for good CL sensitivity (5 keV accelerating voltage, 1 nA probe current, and scanning pixel size of 2 nm with 25 μs dwell time), the rapidly decaying Si-V containing nanodiamonds provide CL images with minimal distortion, whereas the slowly-decaying Eu:Y$_2$O$_3$ nanophosphors (lifetime 2.8 ms [35]) continue to emit optically after the electron beam 'point' excitation has passed, producing streaks in the CL image. Such degraded CL imaging resolution can only be mitigated by scanning more slowly, or by using faster emitters such as Si-V nanodiamonds. Note that it should be possible to further enhance the CL emission rate from Si-V nanodiamonds by increasing the Si and Si-V concentration during CVD growth. For example, $3\times10^{19}$ cm$^{-1}$ Si concentration in diamond has been reported [36]. Advanced photonics techniques, such as directed radiation [28] and plasmonic enhancement [37], may also improve the emission rate from Si-V nanodiamonds, enabling a uniquely bright and non-bleaching CL marker in the near-infrared.

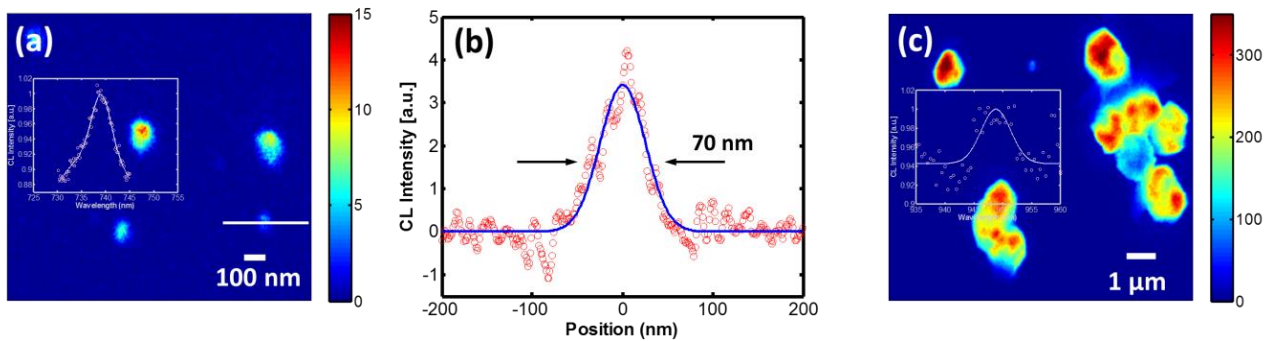

*Figure 3.* CL imaging of different charge states of Si-V color centers in nanodiamonds. (a) With 750±25 nm optical bandpass filter, image of nanodiamond CL produced by Si-V centers in negative charge state. Inset: CL spectrum for (Si-V)⁻. (b) Cross-sectional view of (Si-V)⁻ nanodiamond CL, as marked in (a) by long white bar, with 70 nm FWHM. (c) With 935±25 nm optical bandpass filter, image of micron size diamond CL produced by neutral charge state Si-V centers. Inset: CL spectrum for (Si-V)⁰. Images in (a) and (c) were processed with background subtraction and Gaussian filtering.

For correlative CL and SE microscopy applications, the effective spatial resolution of colored CL markers is determined by nanoparticle size, as the electron beam (diameter <10 nm) induces electron-hole pairs that diffuse over micron-scale distances before recombining, thereby exciting CL approximately uniformly over the nanoparticle volume. The smallest detectable particle size is determined by a combination of factors including CL emitter density and excitation efficiency; efficiency of electron energy deposition into the nanoparticles at a given accelerating voltage; and collection efficiency of the CL photon detection optics. For example, **Figure 3a,b** demonstrate sensitive 2D CL imaging of nanodiamonds as small as 70 nm, for experimental conditions described in the Experimental Section. With optical interference filters switched to transmit CL from the neutral Si-V charge state [(Si-V)⁰, ZPL wavelength ~947 nm] instead of the negative charge state [(Si-V)⁻, ZPL wavelength ~738 nm], the CL brightness greatly decreases, such that the smallest detectable diamond particles are approximately 300 nm in diameter (**Figure 3c**). Successive CL imaging of a sample of ~1 μm Si-V diamond particles gives a ratio of (Si-V)⁻/(Si-V)⁰ CL count rates ≈50, which is of the same order of magnitude as the known ratio of detector quantum efficiencies (≈200) at the two CL wavelengths. This result suggests that the intrinsic CL emission rates of (Si-V)⁻ and (Si-V)⁰, which are proportional to the product of the number of color centers of each charge state present and their radiative decay rates, are similar for the nanodiamonds we synthesized. However, the superior quantum efficiency of most large detection area, low noise, single element optical detectors at shorter wavelengths makes CL imaging of (Si-V)⁻ more practical, and thus it is desirable to tailor nanodiamond synthesis to favor the formation of (Si-V)⁻ centers.

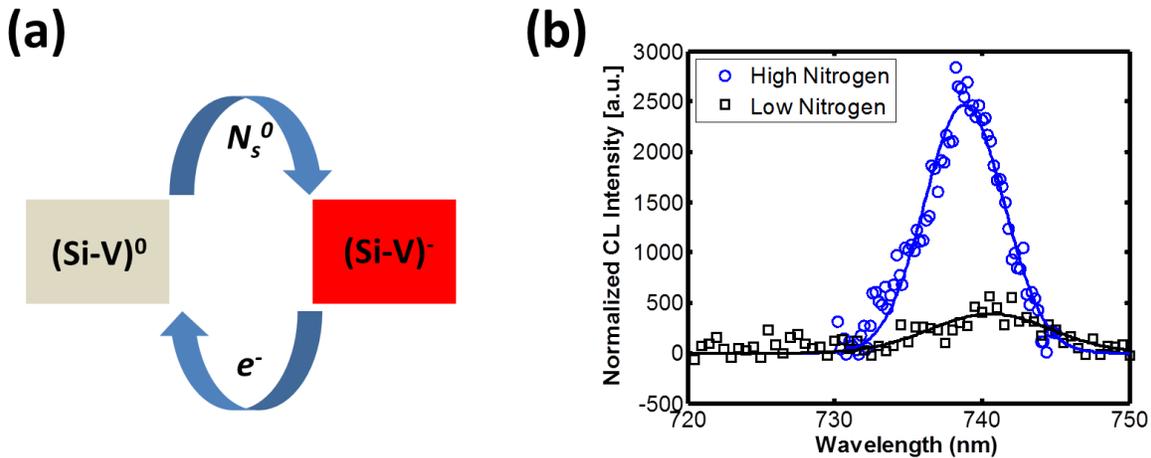

*Figure 4.* Enhanced (Si-V)⁻ CL intensity in nanodiamonds fabricated with elevated nitrogen (N) concentration. (a) Schematic illustration of charge state conversion between (Si-V)⁰ and (Si-V)⁻ because of two competing processes: electron beam ionization of (Si-V)⁻ to (Si-V)⁰; and electron donors from N impurities, which favor (Si-V)⁰ to (Si-V)⁻ conversion. (b) Observed nanodiamond CL spectra shows a

*more prominent (Si-V)⁻ CL line for elevated N concentration. Data are normalized to A-band CL emission peak, and fit to Gaussian function (solid curves).*

We experimentally demonstrated enhancement of (Si-V)⁻ CL intensity in nanodiamonds by introducing nitrogen gas during diamond growth. Isolated nitrogen (N) impurities in diamond are expected to act as electron donors that favor the formation of (Si-V)⁻ centers, and increased CL intensity at 738 nm associated with elevated N concentration has previously been observed in diamond films [38]. **Figure 4a** illustrates two competing charge state conversion processes relevant to CL imaging: (i) electron beam ionization of (Si-V)⁻ to (Si-V)⁰, which is a widely observed phenomenon for vacancy-related color centers in diamond [39][40]; and (ii) increased N impurity concentration within diamond, which enhances the number of available electron donors, such that (Si-V)⁰ is efficiently converted to (Si-V)⁻ [41]. As shown in **Figure 4b**, we observe that CL emission from (Si-V)⁻ at 738 nm is enhanced by a factor of ~6 in nanodiamonds grown with high N concentration compared to those with low N concentration. (See Experimental Section for synthesis details.) We also found that the ratio of (Si-V)⁻ to (Si-V)⁰ CL intensity is higher for nanodiamonds with high N concentration (see **Figure S3**). Further investigation is required to optimize the relative Si and N concentrations for maximal (Si-V)⁻ CL yield, but these measurements clearly illustrate the potential for engineering increased CL signal strength in Si-V nanodiamond biomarkers.

In summary, we performed the first systematic investigation of cathodoluminescence (CL) emission properties of disperse nanodiamonds containing Si-V color centers. Si-V nanodiamonds were synthesized by standard CVD diamond film growth in the presence of silicon, followed by sonication to generate free particles. CL-emitting defects in nanodiamonds are promising as bright, non-bleaching optical markers for correlative CL and SE microscopy, potentially enabling nanoscale imaging of biological structures and simultaneous co-registered localization of multiple molecular species. Our results add Si-V color centers to the set of known CL-emitting defects in nanodiamonds, increasing the potential for wavelength-selective multiplexing, and providing a convenient CL marker in the near-infrared transmission window of biological tissue. We found that the concentration of Si-V color centers in the more readily CL-detectable negative charge state, which emit around 738 nm, can be enriched by synthesizing nanodiamonds with high concentrations of nitrogen electron-donors. Future challenges include increasing the concentration of Si-V centers to optimize CL brightness, and targeted nanodiamond surface functionalization.

## Experimental Section

Nanodiamonds were grown via PECVD (Seki Technotron, Model: AX5010-INT) with conditions of 960 °C, 65 torr, $CH_4:H_2$ ~1:100, microwave power ~ 1050 W. To incorporate more nitrogen ($N_2$) into diamond growth, lower gas pressure (45 Torr) and smaller microwave power (900 W) were intentionally used.[42] These changes not only affect CVD dynamics and nitrogen incorporation efficiency, but also efficiently vary the N/C ratio when considering the air leak rate for our chamber. Nanodiamond samples were prepared for TEM observation as follows: (i) sonication of nanodiamonds on a silicon wafer for more than 10 minutes; (ii) drop-cast of 5 μL of the sonicated solution onto a TEM finder grid using a micropipette; and (iii) manipulation of specific nanodiamonds on the TEM grid using a nano-manipulator (Omniprobe 200) integrated inside a Zeiss NVision 40 Dual-Beam FIB and SEM. See **Figure S1**. Nanophosphor Eu:$Y_2O_3$ powders (Boston Applied Technologies) were ultrasonicated with a probe sonicator (Branson 450) at 120 W average power for 30 minutes, and then fractionated by repeated centrifugation to remove agglomerates.

CL properties were investigated with a field emission SEM (JEOL JSM-7001F) fitted with a spectrally-selective, PMT-based CL detection system (KE Developments Centaurus CL detector). A PMT (ET Enterprises 9798B) was used for its extended response into the infrared. Bandpass optical filters were employed at 725-775 nm (Chroma HQ 750/40m) for $(Si-V)^-$ CL measurements, and at 910-960 (Chroma HQ 935/40m) nm for $(Si-V)^0$ CL studies. A home-built system was implemented for CL spectral measurements. Multi-mode fiber (Thorlabs FT200UMT) brought the CL signal from inside the SEM vacuum chamber and into a monochromator (Princeton Instruments, Acton SP-2358 with a grating of 300 g/mm and 500 nm blaze). At the spectrometer output, the CL signal was again fed into a multi-mode fiber and then detected by an APD single photon counting module (Perkin Elmer SPCM-AQRH-13-FC). The typical spectral step size was 0.5 nm, with a 1 sec integration time for each step.

TEM measurements were performed with a JEOL 2100 TEM, with a LaB6 gun running at 200 kV. For samples in solution, PL spectra were measured with a Tecan Safire$^2$ plate reader. For dry samples, PL spectra were measured in a home-built confocal microscope (532 nm laser excitation). This confocal microscope was also used for all Raman spectral measurements.

## Acknowledgements

We acknowledge the financial support from the NSF and DARPA. Dr Aharonovich is the recipient of an Australian research council discovery early career research award (project number DE130100592).We thank Mary Chessey and Jean-Christophe Jaskula for helpful technical assistance.

# Supporting Information

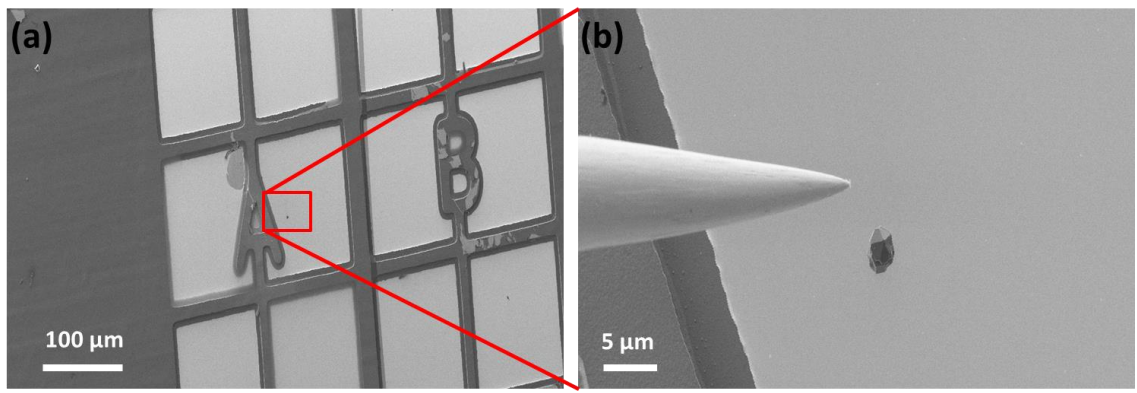

*Figure S1.* Transfer of Si-V nanodiamonds (NDs) from silica substrate to TEM grid. (a) Released ND sitting on top of patterned TEM finder grid. (b) Zoomed in image showing ND along with nanomanipulator tip (Zeiss NVision 40 FIB equipped with Omniprobe nanomanipulator).

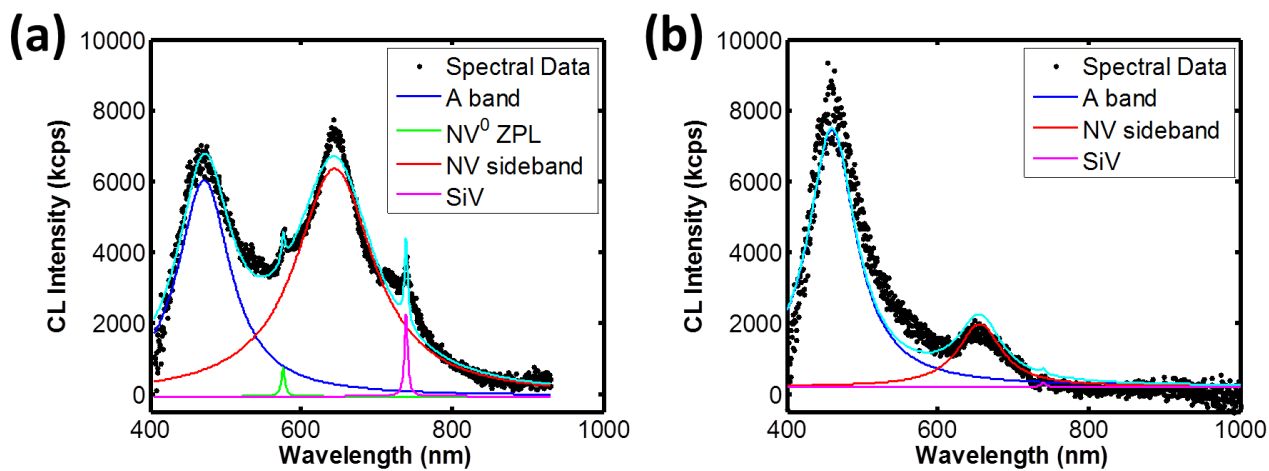

*Figure S2.* CL spectra from Si-V nanodiamond samples with (a) high nitrogen concentration, and (b) low nitrogen concentration. Black dots indicate CL experimental data. Solid color lines are from fits of data to model CL spectra for different color centers and defects. The high N concentration sample shows a sharp peak for (Si-V)⁻ CL, which is nearly absent in the low N concentration sample. NV CL signals are also greatly enhanced in the high N concentration sample.

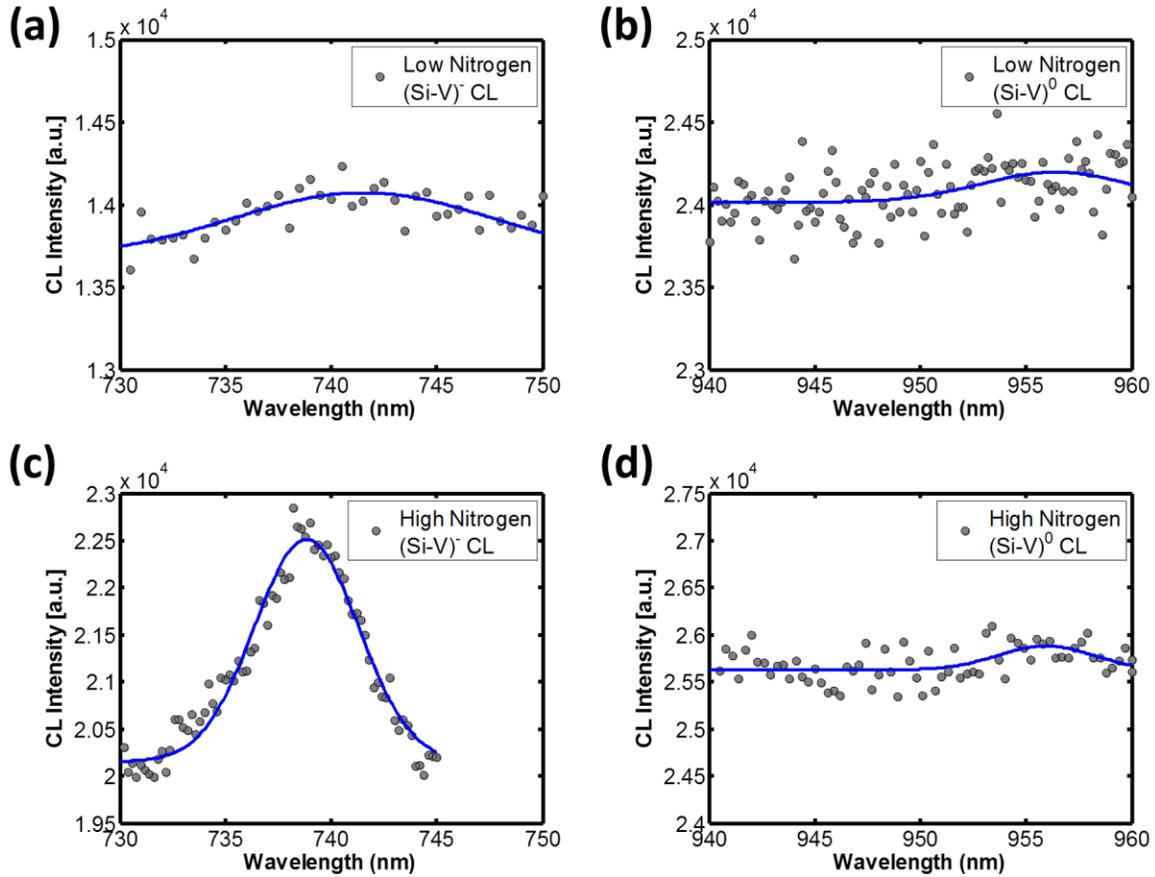

*Figure S3*. *Comparisons of measured (Si-V)⁻ and (Si-V)⁰ CL spectra. (a) and (b): low N concentration nanodiamond CL spectra for (Si-V)⁻ and (Si-V)⁰, respectively. (c) and (d): high N concentration nanodiamond CL spectra for (Si-V)⁻ and (Si-V)⁰, respectively. Integration time for each data point is 5 s for (a) and (c) and 10 s for (b) and (d). From these CL spectra we estimate the (Si-V)⁻ to (Si-V)⁰ ratio ≈0.9 for the low N concentration sample and ≈4.9 for the high N concentration sample (corrected for detector quantum efficiency and spectrometer grating efficiency), which is consistent with data from images with different optical filters (see main text). Solid lines are fits to a single Gaussian function.*